\documentclass[preprint]{aastex7} 
\newcommand{\alf}{Alfv\'en}
\newcommand{\ha}{H$\alpha$}

\newcommand{\sm}{$\sim$}
\newcommand{\kms}{km~s$^{-1}$}
\newcommand{\sdo}{\textit{SDO}}

\newcommand{\iris}{\textit{IRIS}}

\newcommand{\hri}{EUI/HRI$_{\rm EUV}$}

\newcommand{\lee}[1]{{\color{black}#1}}

\shorttitle{Tiny Jets detected by Solar Orbiter and Big Bear Solar Observatory}
\shortauthors{Lee et al.} 

\begin{document}
%\title{Tiny Jets observed with by Solar Orbiter and Big Bear Solar Observatory and Puzzles on Magnetic Reconnection}
\title{Fine Structures of Tiny Quiet Sun Jets Observed by Solar Orbiter and Big Bear Solar Observatory}
%The First Joint Observations of EUV Jets and spicules with SolO and BBSO
\author[0000-0002-5865-7924]{Jeongwoo Lee}
\affiliation{Institute for Space Weather Sciences, 
            New Jersey Institute of Technology,
            University Heights, Newark, NJ 07102, USA}
\affiliation{Center for Solar-Terrestrial Research, 
            New Jersey Institute of Technology, 
            University Heights, Newark, NJ 07102, USA}
\affiliation{Big Bear Solar Observatory,
            New Jersey Institute of Technology, 
            40386 North Shore Lane, Big Bear City, CA 92314, USA}
\email{leej@njit.edu}
\correspondingauthor{Jeongwoo Lee}
\email{leej@njit.edu}

\author[]{Dana Longcope}
\affiliation{Dept. of Physics, Montana State University, Bozeman, MT 59717, USA}
\email{dana@physics.montana.edu}

\author[0009-0008-6476-054X]{Junmu Youn}
\affiliation{School of Space Research, Kyung Hee University, Yongin 17104, Republic of Korea}
\email{jmyoun@khu.ac.kr}

\author[0000-0001-7620-362X]{Navdeep K. Panesar}
\affiliation{Lockheed Martin Solar and Astrophysics Laboratory, 3251 Hanover Street, Building 203, Palo Alto, CA 94306, USA}
\affiliation{SETI Institute, 339 Bernardo Ave, Mountain View, CA 94043, USA}
\email{panesar@lmsal.com}

\author[0000-0001-9049-0653]{Nengyi Huang}
\affiliation{Institute for Space Weather Sciences, 
            New Jersey Institute of Technology,
            University Heights, Newark, NJ 07102, USA}
\affiliation{Center for Solar-Terrestrial Research, 
            New Jersey Institute of Technology, 
            University Heights, Newark, NJ 07102, USA}
\affiliation{Big Bear Solar Observatory,
            New Jersey Institute of Technology, 
            40386 North Shore Lane, Big Bear City, CA 92314, USA}
\email{Nengyi.Huang@njit.edu}

\author[0000-0002-5233-565X]{Haimin Wang}
\affiliation{Institute for Space Weather Sciences, 
            New Jersey Institute of Technology,
            University Heights, Newark, NJ 07102, USA}
\affiliation{Center for Solar-Terrestrial Research, 
            New Jersey Institute of Technology, 
            University Heights, Newark, NJ 07102, USA}
\affiliation{Big Bear Solar Observatory,
            New Jersey Institute of Technology, 
            40386 North Shore Lane, Big Bear City, CA 92314, USA}
\email{haimin.wang@njit.edu}

\begin{abstract} 
We present the first joint high-resolution observations of small-scale EUV jets using Solar Orbiter(SolO)'s Extreme Ultraviolet Imager and High Resolution Imager (\hri) and \ha\ imaging from the Visible Imaging Spectrometer (VIS) installed on the 1.6~m Goode Solar Telescope (GST) at the Big Bear Solar Observatory (BBSO). These jets occurred on 2022-10-29 around 19:10 UT in a quiet Sun region and their main axis aligns with the overarching magnetic structure traced by a cluster of spicules. However, they develop a helical morphology, while the {\ha} spicules maintain straight, linear trajectories elsewhere.  Alongside the spicules, thin, elongated red- and blue-shifted \ha\ features appear to envelope the EUV jets, which we tentatively call sheath flows. The EUI jet moving upward at speed of \sm110 {\kms} is joined by strong \ha\ red-shift \sm20 {\kms} to form the bidirectional outflows lasting \sm2 min. Using AI-assisted differential emission measure (DEM) analysis of SolO's Full Sun Imager (FSI) we derived total energy of the EUV jet as \sm$1.9 \times 10^{26}$ erg with 87\% in thermal energy and 13\% in kinetic energy. The parameters and morphology of this small-scale EUV jet are interpreted based on a thin flux tube model that predicts {\alf}ic waves driven by impulsive interchange reconnection localized as narrowly as \sm1.6 Mm with magnetic flux of \sm$5.4\times 10^{17}$ Mx, belonging to the smallest magnetic features in the quiet Sun. This detection of intricate corona--chromospheric coupling highlights the power of high-resolution imaging in unraveling the mechanisms behind small-scale solar ejections across atmospheric layers. 
\end{abstract}

%% The AAS Journals now uses Unified Astronomy Thesaurus concepts:
%% https://astrothesaurus.org

%%%%%%%%%%%%%%%%%%%%%%%%%%%%%%%%%%%%%%%%%%%%%%%%%%%%%%%%%%%%%%%%%%%
\section{Introduction}\label{sec:intro}
Small-scale ejections/eruptions (SSEs) in the low solar atmosphere are believed to play an important role in the energy balance, mass loading, and fine structuring of the solar corona as well as mass transport into the solar wind \citep{Lee_2022, Sterling2024, Bizien2025A&A...694A.181B}.  
Many of these small-scale events are observed in a variety of forms of jets, which may provide the upward flux of mass, momentum, and energy necessary for the observed heating and flows \citep{Panesar_2023, ShiF2024}. However, the connections among the different types of SSEs, especially those in different layers of the atmosphere, remain obscure \citep{ShiF2024}. Improved physical understanding of the formation of and the mechanisms behind the jetting phenomenon is fundamentally important within the broad field of Heliophysics \citep{Shen_2021}. Such works could help synthesize the disparate observations and theories of SSEs into a more cohesive, coherent framework.

SSEs are characterized by well-collimated ejecta apparently flowing along preexisting magnetic-field lines. 
First discovered in \textit{Skylab} He~{\sc ii} 304~\AA\ images as cool (\sm8~$\times$10$^4$~K) plasma ejections in coronal holes \citep{Bohlin1975_macrospics}, macrospicules appear to be giant spicules or small surges \citep{beckers77,Tandberg-Hanssen77}. They extend 7000--40,000~km above the limb, with a rising velocity of 10--150~\kms\ and a lifetime of 3--45 minutes \citep{Dere1989_macrospic,Karovska1994_macrospic}. Using data from Big Bear Solar Observatory (BBSO), H$\alpha$ macrospicules were found to be associated with EUV macrospicules and X-ray bright-point flares \citep{Moore1977}. Later work \citep{Yamauchi2004,Yamauchi2005} showed that macrospicules exhibit two different forms: an erupting loop containing a mini-filament (see below), and a single-column spiked jet. 
The relationship between macrospicules and other SSEs is unclear, particularly since previous studies were focused on limb events for which no magnetic-field measurements were available. 
 
Higher in the atmosphere, observations in UV spectral lines reveal the presence of similar kinds of ejections in the transition region (TR). Impulsive bidirectional and unidirectional high-speed flows denoted  ``explosive events'' and ``coronal bullets'' were first detected in UV spectra by the \lee{High Resolution Telescope and Spectrograph \citep[HRTS; ][]{Cook1983}} rocket-borne instrument \citep{Brueckner1982}, and attributed to magnetic reconnection in the network \citep{Dere1989}. 
Analyzing Si {\sc iv} 1402.77 \AA\ line data from \lee{Interface Region Imaging Spectrograph} \citep[{\iris};][]{DePontieu.IRIS.2014SoPh..289.2733D}, a prevalence of intermittent small-scale TR jets with speeds of 80--250~\kms\ originating from the narrow bright network lanes was found \citep{Tian+etal.BD.2014ApJ...790L..29T}. These jets have lifetimes of 20--80 s and widths of $\le$300 km, and are rooted in small-scale bright regions, often preceded by footpoint brightenings and accompanied by transverse waves with 20~\kms~amplitudes. Many TR jets reach temperatures of at least 10$^5$~K and constitute an important element of the TR structure. As with Type II spicules and macrospicules, we propose to explore the relationship between TR jets and other SSEs.  

Coronal jets are commonly found in coronal holes, quiet-Sun regions, and the peripheries of active regions. First detected in soft X-ray images by Yohkoh \citep[e.g.,][]{shibata92}, coronal jets have since been observed in the EUV, hard X-rays, white light (via coronagraphs), and microwaves, with speeds of \sm100--400~\kms\  and sizes of \sm5--500~Mm \citep{Raouafi2016}. \lee{Small-scale jetting activity, also called jetlets \citep{raouafi14,Panesar2018}, has received attention for their possible connection to solar wind transients, including switchbacks \citep{Raouafi2023ApJ...945...28R}.}
Recent studies, using Solar Orbiter's data (SolO; \citealt{Muller_2020}), have turned to even smaller-scale ejections, campfires, probably the smallest class of EUV jets observed in the corona \citep{Berghmans_2021, Panesar_2021, Panesar_2023}, providing unprecedented insights into these fine-scale phenomena.

A new opportunity for studying fine-scale SSEs is available with the SolO  in collaboration with 
\lee{Goode Solar Telescope (GST) at Big Bear Solar Observatory (BBSO)}.  SolO includes the Extreme Ultraviolet Imager (EUI), which offers high-resolution images through the High Resolution Imager, HRI$_{\rm EUV}$ at 174 \AA\ with a pixel resolution of $<$0.5\arcsec\ \citep{Rochus_2020, Berghmans_2021}, which corresponds to 100 km on the Sun during the perihelion observations \citep{Rochus_2020}.
\lee{\citet{Chitta2023Sci...381..867C} 
reported so-called picoflare jets on scales of a few hundred kms and speeds of \sm100 \kms\ detected by the {\hri}, which would be one of the smallest-scale jets. 
\citet{Petrova2024A&A...687A..13P} found a tiny helical flux rope at the propagation speeds of 136--160 \kms\ using the Spectral Imaging of the Coronal Environment (SPICE)  instrument and Polarimetric and Helioseismic Imager (PHI) onboard SolO.
}
On the ground, the 1.6~m, high-order \lee{ adaptive optics} (AO) equipped GST at BBSO can achieve resolution as high as 0.1\arcsec--0.2\arcsec with the pixel size of 0.029\arcsec, ideal for studying small-scale physical processes in the photosphere and chromosphere \citep{goode10}.  Time series of \ha\ imaging spectroscopy with multiple wavelength points provide the Fabry-P\'erot-based Visible Imaging Spectrometer (VIS) of high image quality  \citep[e.g.][]{Samanta2019}.
Such SolO--BBSO joint observations allow us to address %on allows us to make cross-comparison of the photospheric, chromospheric, and coronal part of jets, and 
the key issue of \lee{SSEs}: what dictates the structure and dynamics of the SSEs and what role the magnetic reconnection has in SSEs 
in the photosphere, chromosphere, and the corona.

\section{Data and Analyses}\label{data}

\begin{figure}[ht]
    \centering
   \includegraphics[width=\textwidth]{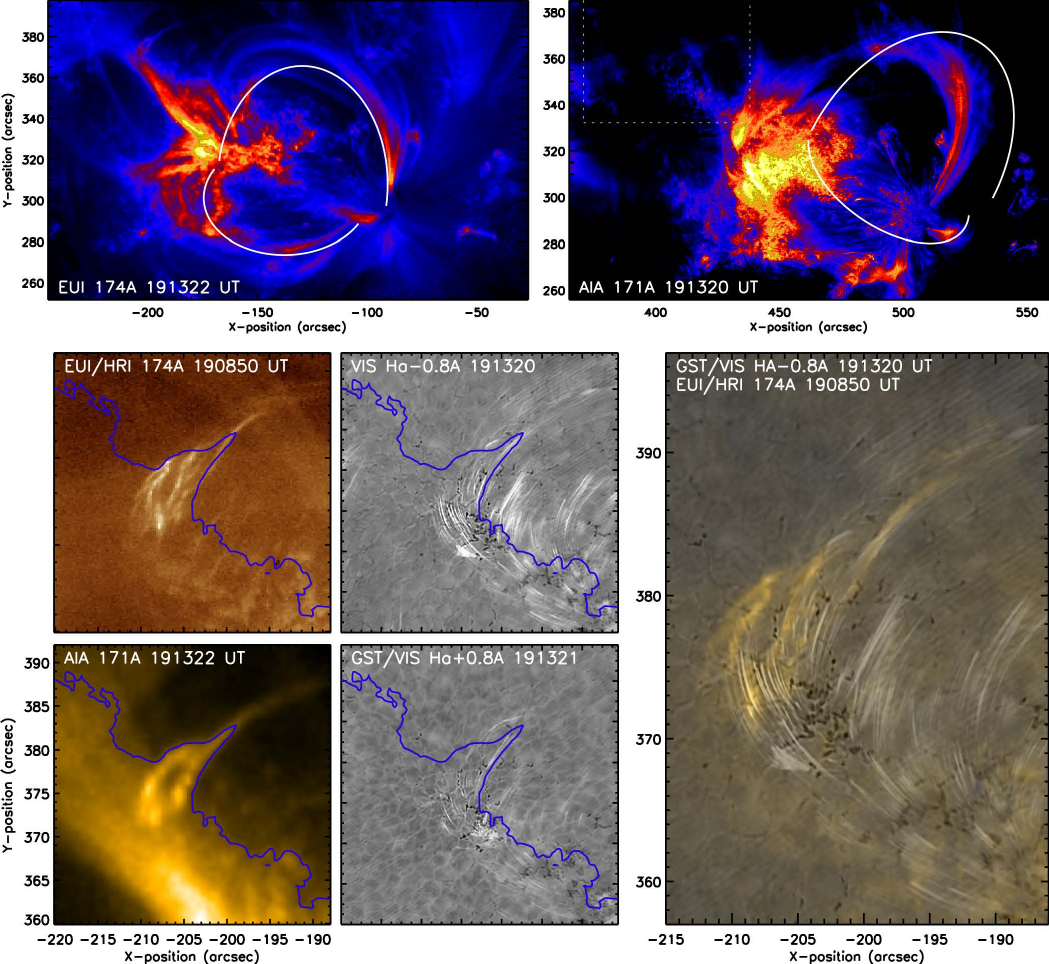}
\caption{Projection effects of the quiet-region EUV jet for multiple instruments. ($a$) SDO/AIA 171~\AA\  with a small jet encloseed by the dotted box.
($b$) SolO's \hri\ 174 {\AA} of the same region.
The three-color half circles are meant to show the projection effect. Model \lee{loops} are constructed in the disk center and projected back to the two locations.
Other images in the small FOV around the jets shown in ($c$--$e$) include a \lee{tilt}-adjusted \hri\ 174 {\AA} ($c$), the \sdo/AIA 171 {\AA} ($d$) and \ha\ wing $\pm$0.8 \AA\ images ($e,f$). As a reference, we overlay the outline of the AIA 171 \AA\ profile (blue contour from $d$) on other images ($c, e, f$).
A composite image of the \ha\ blue wing image (gray-scale) and \hri\ in yellow color shades ($g$) shows different morphologies.
The observation time of AIA and  GST is delayed by 4.5 min from that of {\hri}.
}
\label{f1}
\end{figure}

On 2022-10-29, the GST was pointing to a quiet Sun region (201$''$E, 356$''$N), which was also within the field of view (FOV) of SolO's {\hri}. During the coordinated observation, \hri\ detected two small-scale jets in quick succession at 19:07:30 UT and at 19:13:30 UT at the position of SolO. 
SolO was located  at 0.46 AU to result in the photon arrival time difference of 4.5 min with respect to SDO and BBSO. SolO had angular separation from the SDO by 37$^{\rm o}$.9 so that the target in the east quadrant in AIA view appears in the west quadrant in the SolO's view (Figure \ref{f1}). 
%Tiny EUV jets (marked with a box) emerged suddenly from the dark corona, which was caught by SolO's {\hri}. 
We were unable to find, in the SolO's SPICE data, significant Doppler signals at the jet location, likely due to either weak emission or the large inclination angle of the jets relative to the line of sight (LOS). \ha\ pseudo-Dopplergrams are constructed from the GST/VIS images at five wavelength points, \ha$\pm$0.8 \AA\, \ha$\pm$0.4 \AA\ and the line center.
We used the center-of-gravity (COG) method, which utilizes the residual intensity profile, the difference between the line profile and the reference profile. 
Depending on which reference profile is used, two types of Doppler speed may result. Use of the ambient continuum intensity as the reference \citep{Uitenbroek2003} gives  higher weights to the line core, resulting in low speeds. If we use the mean spectrum of the whole FOV \citep{Rouppe+etal2009ApJ...705..272R}, more weight is given to the wing enhancement resulting in higher speeds. The latter method was used in this study.

\subsection{SolO/EUI, SDO/AIA and GST/VIS Images}

Figure \ref{f1} presents SDO/AIA 171 \AA\ image ($a$) and \hri\ image in larger FOV ($b$) including NOAA AR 13133 itself in top panels, where our target jet is marked with the white box. Zoom-in views of the jet seen by three different instruments are plotted in the bottom four panels ($c$--$f$) along with an overlay of \ha\ and EUV images in ($g$).
The three colored curves and the warped rectangles mark NOAA AR 13133 are not the present target but plotted to check the different projection effects on both images. 
They are coordinate-transformed from the half circles tilted at three angles and a straight rectangle in the disk center to the position of AIA and that of \hri\ to demonstrate the different orientations of the two loops in the SDO and SolO images.
Our target is a tiny EUV jet that emerged suddenly from the dark corona around 19:10 UT within the region marked with the box located north off the AR. 
The main parts of this jet are obscured by an EUV loop in the foreground in SDO/AIA  ($a$), but could directly be detected from the different perspective of SolO  ($b$).
For the direct comparison with AIA images, we adjusted the tilt angle of the \hri\ 174 {\AA} according to the above practice of the projection on different locations on the disk ($c$).
%If we forget the fine structure inside, the jets observed in AIA 171 \AA\ and EUI/HRI 174 \AA\ (white box in panel b) exhibit similar height of the jets with different footpoints separation. It is that the jets in SolO's view are positioned farther from the disk center than SDO and BBSO, and the \hri\ images experience more foreshortening. To compensate this, we stretched the \hri\ images along the solar X-direction to align them with AIA and GST data, matching their spatial scales. This alignment technique was applied in Figures 2 and 4, and there we adopted AIA coordinates for \hri\ images. 
The blue contours outlining profiles of the AIA image are taken from ($d$) and copied to other panels ($c,e,f$).
The \hri\ image reveals the fine-scale helical pattern of EUV bright strands giving an impression of twisted field lines or flux rope ($c$). The jet's fine structure is less clearly visible in the AIA 171 \AA\ image ($d$) although AIA 171 \AA\ images usually look similar to the EUI 174 \AA\ images \citep[e.g.,][]{Mandal2023}. 
%For instance, The helical structure implied by the three parts: A--C is not clearly identified in AIA is repeatedly seen in AIA, but B and C are overlapped to be seen as one strand. As a result, with the AIA 171 \AA\ alone it is not so clear whether it is.
The different appearance can partly be due to the lower spatial resolution of AIA or the accidental blockage of the jets by a foreground EUV bright structure in AIA's FOV might be another factor. 
We must also note that the EUV structures appear different shapes depending on the observer’s different perspective, as previous DEM studies comparing SDO/AIA and Solar Orbiter/FSI data have demonstrated \citep{youn2025can}.

The GST/VIS's inverted {\ha} blue wing ($e$) and red wing ($f$) images exhibit the spicules in straight trajectories without any helical structure of the EUV jets. A composite image of the \ha\ blue wing image (gray-scale) and \hri\ in  yellow color shade ($g$) shows an obvious difference  that the \ha\ spicules move along straight trajectories and the \hri\ 174 \AA\ structures are twisted. 
In addition, there are differences in spatio-temporal distribution. The EUV jets occurred in a small area (\sm3 Mm) and lasted only \sm2 min in the region where otherwise is EUV dark, while the spicules are numerous and constantly occurring elsewhere at an enormously different rate.
All of these differences should arise from the intrinsic properties in both radiations and are hardly due to either the projection effect or resolution.
As an additional note, no jet bright point \citep[JBP;][]{Sterling2016} is found, implying that the EUV jet only occurred up in the corona without leaving traces in the lower atmosphere. In that sense, this jet is similar to some of the Hi-C jets \citep{Panesar2019} that occurred at the edges of the magnetic network lanes.

\subsection{EUV Jets and Spicules}

\begin{figure}[ht]
    \centering
  \includegraphics[width=\textwidth]{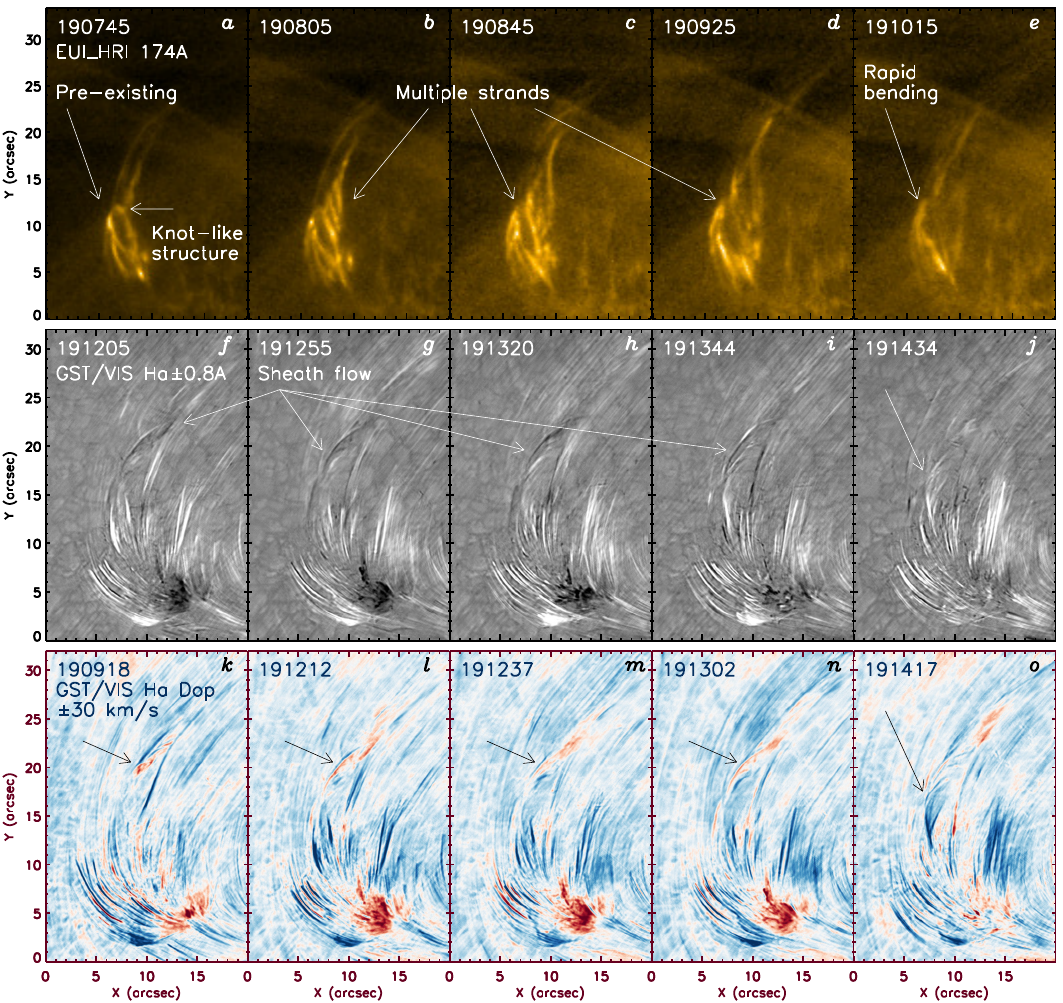}
\caption{The evolution of \hri\ jet structure at 174~\AA\  (top),  difference between H$\alpha \pm$0.8 {\AA} wing images (middle), and pseudo-Dopplergram from 5 wavelength point GST/VIS images (bottom). 
They show pre-existing strand ($a$), knot-like structure ($b$), multiple stripes ($b$--$d$) and a rapidly turning or bending structure ($e, j, o$) as well as sheath flows (arrows in $f$--$i$, $k$--$n$). The Dopplergrams show strong redshift components concentrated under the EUV jets. 
}
\label{fig.so_jet}
%\vspace{-3.0mm}
\end{figure}

In Figure \ref{fig.so_jet}
we make one-to-one comparisons of the EUV jet with those of \ha\ spicules during their evolutions.
These jets are highly inclined, and  we nonetheless plot them with their axis up for visual convenience.
Top panels show a sequence of the \hri\ images at the time of the first jetting: 19:07 UT - 19:10 UT alongside the GST \ha\ far wing difference images in the middle, and \ha\ Dopplergrams in the bottom.
Initially (before 19:07 UT in SolO time) there was a faint structure looking like a weakly curved flux tube, which develops a brightening in the middle of the tube to form a wedge like structure and further evolves into a knot-like structure ($a$). This is suggestive of interchange reconnection in appearance. 
The twisted structure with a single strand develops into multiple strands ($b$) and three prominent stripes at the maximum intensity ($c$). Their separation increases as they propagate away ($d$) and finally disappear ($e$). We also see the points where the jet's guiding field lines rapidly turn around ($d$); it is either the curved jet structure or some other background structure. In about 4 min, the jet diminished and a faint rapidly bending structure appears in the \lee{flank} of the EUV jet ($e$).

%%The stark differences between the EUV jets and \ha\ spicules in terms of morphology and numbers. Whether the evolving complicated EUV intensity stripe pattern (d) should represents either instantly forming flux ropes or {\alf} wave train is an important issue to investigate.

The middle panels show the red wing to blue wing difference \ha$\pm$0.8 \AA\ images at the corresponding times taking into account the relative time delay between \hri\ and GST.
One similar feature of the \ha\ images with the \hri\ images is that the strand in ($f$) well coincides with the pre-exiting EUV structure ($a$).   
%On the tip, it is is sharply curved ($f$--$i$), which seemingly coincides with the faint rapidly bending structure in the frank of the EUV jet ($e$). It looks either like a rapidly bent flux rope or an overlapping background feature. This EUV feature is not well visible in AIA but is in \hri\ image owing to its higher resolution.
The multiple EUV stripe pattern is not exactly reproduced in the \ha\ images, but an alternating thin blue- and red-wing feature as thin as 200 km is found near the boundary of the EUV jets ($f$--$i$). Given their spatial association with the jet, we tentatively call it `sheath flow,' which may be cold plasma enveloping the hot EUV jet stream. The sheath flow are aligned toward the EUV jet spire with their characteristics different from typical spicules. 
Later the sheath flow shrinks down to a lower height ($j$), although it does not exactly coincide with the rapidly bending EUV jet structure ($e$). They are therefore not the same phenomena.

\lee{The sheath flow is a new phenomenon requiring further interpretation. In this simple picture, the EUV-bright and dark regions correspond to the corona and chromosphere, respectively. The sudden appearance of an EUV jet within a dark region is thus regarded as the formation of a localized corona embedded within the chromosphere rather than above the chromosphere. This structure likely introduces a thin interface between the hot jet and the surrounding chromospheric plasma that forms due to insufficient time for full exchange of particles and energy between them. Nearby spicular flows will be sliding along the sheath, instead of penetrating the jet region, like a tangential flow along an impermeable wall, and thus called sheath flow.}
Other up-rising spicules are also in the straight trajectory in contrast with the helical EUV jet structure. We should also note that the \ha\ spicules are constantly regenerating, whereas the EUV helical structure last over a short period \sm2 min.  
%The large temperature difference between EUV jets and \ha\ spicules may explain why the \ha\ spicules run around but not inside the EUV jets.

The Dopplergrams in the bottom panels show the sheath flows more clearly (arrows in $k$--$n$), which could imply torsional motion of cool plasma around the EUV jets.
The spray feature in the \ha\ wing difference image ($j$) appears as a blueshift component in the Dopplergram ($o$). The \ha\ spray feature does not exactly coincide with the rapidly bending EUV structure ($e$), and may be a sheath flow too.
Otherwise the blue-shift component  dominates ($f$--$h$), and those in thin and long structures are upward spicules. 
The red-shifted \ha\ components may also be spicules directed downward, but are timely correlated with the EUV jets  unlike other normal spicules, and are highly concentrated in the presumed location of the EUV jet footpoint. We suggest that they could rather be the downward outflow from the reconnecting X-point associated with the EUV jets. Their speeds reach \sm20 \kms\ and join the EUV jets in the opposite direction. The \ha\ redshift is therefore an important component that connects the EUV coronal jets down to the photospheric magnetic fields.
%, since the EUV jets themselves do not extend down to the photosphere. 

%The origin of the evolving complex pattern of the EUV jets remains as the key question to investigate. The spicules appearing as blueshifts shows a morphological  difference between EUV jets and \ha\ spicules, and in location EUV jets bundle does not allow spicules to get in. Conversely, spicules occur elsewhere in regions of no EUV counterparts, e.g., the EUV group d vs. {\ha} group h). Actually EUV jets (h) are largely missing the majority of {\ha} spicules elsewhere (the issue raised by Lee et al. 2025, Nivied et al. 2022). Note also that numerous {\ha} spicules are steadily replenished whereas EUV jets work only for a limited time.

\subsection{Dynamics of EUV Jets}

\begin{figure}[ht]
    \centering
    \includegraphics[width=\textwidth]{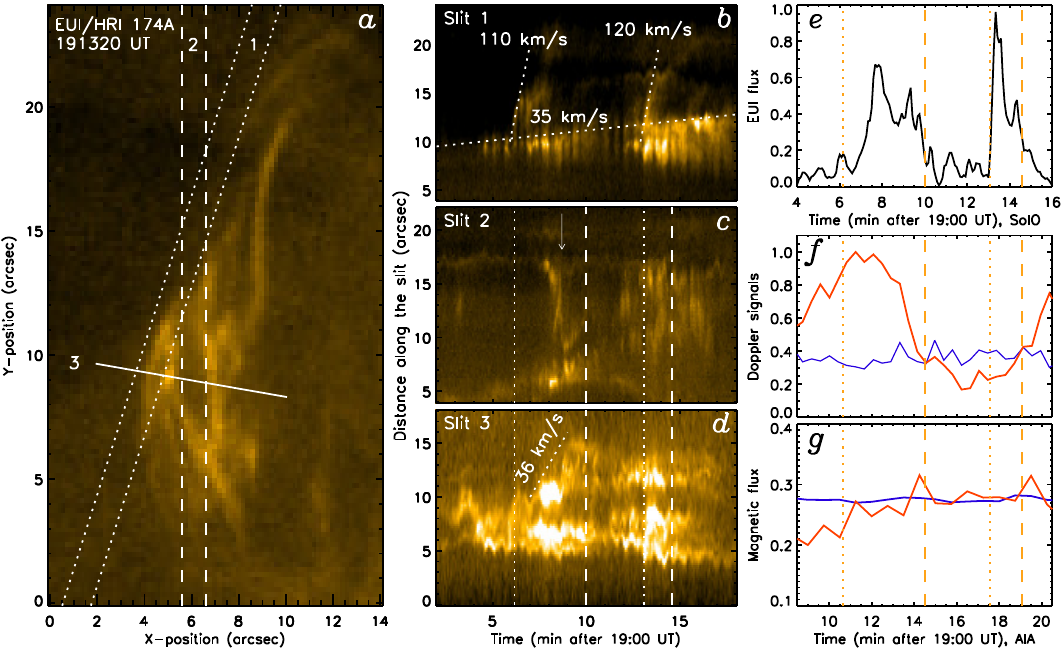}  
\caption{EUV TD maps and time profiles of EUV flux, \ha\ Doppler motions and magnetic flux.
(a) An \hri\ image with slit positions set for the TD maps.
\lee{(b--d)} TD maps for EUI along the slits \lee{\#1--3} show
upward speeds of 110--120 {\kms}, and the evolving helical structure (white arrow). 
Time profiles of \lee{(e) the EUV flux,  (f) {\ha} Dopplershift fluxes, and (g) magnetic fluxes. In (g), the red (blue) curve corresponds to the positive (negative) flux. In (c--g) the vertical guides lines indicate the start (dotted line) and the end (dashed) times of the two jets.}
Animation of the panel (a) from 19:00:00 UT to 19:21:40 UT  is available.}
\label{eui_td}
\end{figure}

We focus on investigating the dynamics of the EUV jets using time-distance (TD) maps, which provide estimates of transverse motions. 
Figure \ref{eui_td} presents TD maps along four slits and related time profiles.
Slit 1, aligned with the most persistent structure allows an easy measurement of the linear upward motion: the first (19:06 UT) and second jets (19:13 UT) exhibit upward speeds of \sm110 {\kms} with a baseline increasing more slowly at \sm35 {\kms}.
However this slope can be underestimated due to the finite inclination angle. 
Therefore, the upward EUV jets and the downward \ha\ materials have enormously different speeds.
Slit 2 set parallel to the jet axis and passing through the first jet finds 
a fast twisting feature at (19:08 UT). 
%Slit 3 is also parallel to Slit 2 but passing through the brightest point of the second jet to find an equally fast rise of the jet, but at a delayed time. They together yield impression of rotating twist.
Slit 3, positioned orthogonally across the jet axis, detects the width of the jet and shows the pronounced expansion occurring primarily during the first jet. The jets expands in width from 2 Mm to 5 Mm for the 3 min period, after which the structure remains stable in size through the second jet.

In the rightmost column, 
we compare  time profiles of the EUI 174 \AA\ flux, the area-integrated \ha\ Doppler signal, and the magnetic flux with each other. 
Due to the SolO--AIA position difference,
The EUI 174 \AA\ flux peaks at 19:07:30 UT and 19:13:30 UT appear at  the SDO position, with the photon arrival time delay by 4.5 min.
During the first peak, the EUV flux peak coincides with that of the area-integrated \ha\ redshift signal although the redshift flux tends to precede the EUV flux enhancement. 
This correlation between EUV jets --\ha\ redshifts is missing for the second EUV jet. Another difference from the first jets is that the EUV bright strands in the second jets are structured like straight flux tubes aligned with the \ha\ spicules.
The magnetic fluxes measured from the HMI magnetograms (see Figure \ref{f_mag}) shows that the negative-polarity magnetic flux remains largely stable, while the parasitic positive flux steadily increases after its emergence covering the period of the first jets. No such trend is seen for the second EUV peak. Therefore the first and the second EUV jets differ from each other in morphology and correlation with the \ha\ Doppler signals and magnetic flux, implying different driving mechanisms for the two EUV jetting events.

\subsection{Magnetic Fields Around the EUV Jets}

\begin{figure}[ht]
\centering
\plotone{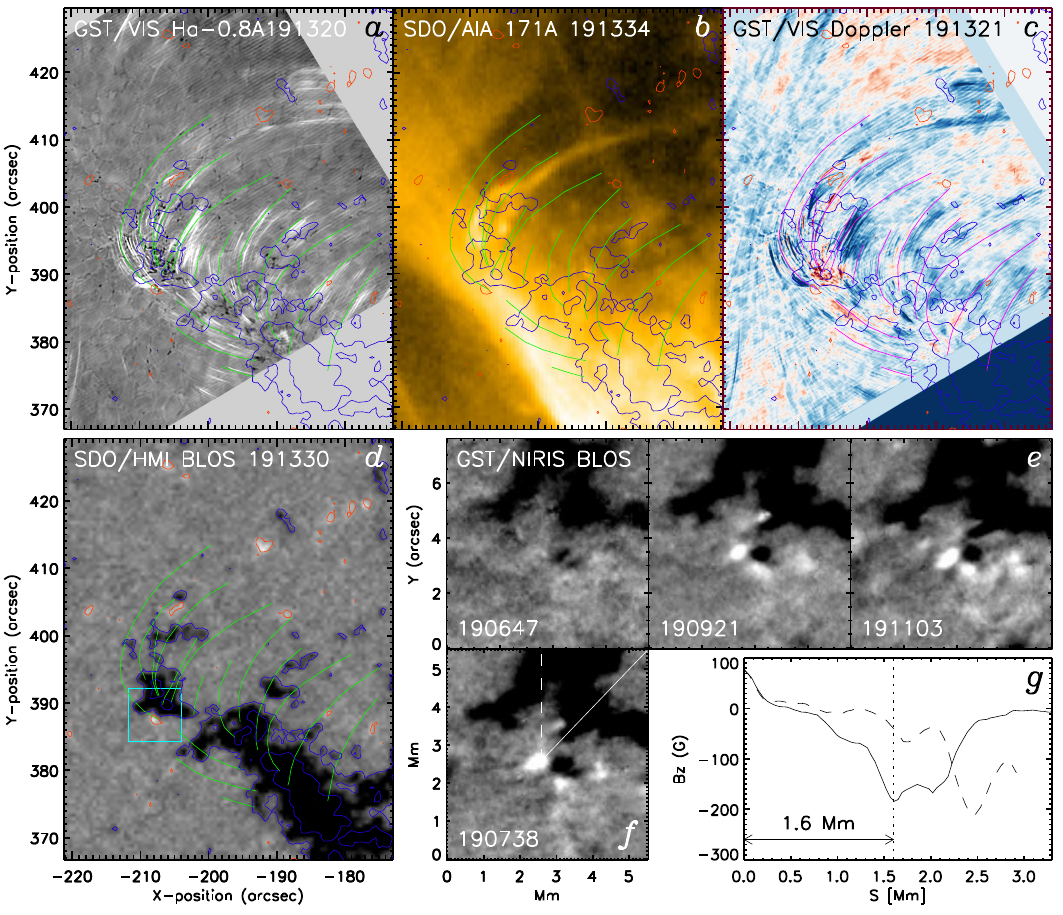}
\caption{\lee{Correspondence between the magnetic fields, spicules, and the EUV jet.} 
(a) The \ha\ spicule trajectories readouts (green lines) from the GST/VIS $-$0.8 \AA\ image are taken as proxy for the magnetic field lines, and copied to (b-d).
(b) \sdo/AIA 171 \AA\ image with the LOS magnetogram (contours).
\lee{(c) Same as (a) over the the Dopplergram with the blue/red colors representing blueshift and redshits. The field lines are colored magenta in this panel. } 
(d) An \sdo/HMI  LOS magnetogram with the contours at the levels of $-$40 G (blue) and $+30$ G (red). These contours are over-plotted in panels (a--c) as well. 
(e) The three panels are GST/NIRIS LOS magnetic fields in the small FOV (cyan box in d) in units of arcsec. 
%The display of the grayscale images are confined to $\pm$30 G. 
(f) NIRIS magnetogram in the same FOV at another time in units of Mm. Two guide lines are the slits for scanning the 1-D magnetic field distributions.
(g) Distribution of the vertical magnetic field, $B_z$, read along the two slits. 
Animation shows the panels (a) and (c) from 19:06:16 UT to 19:20:55 UT.}
\label{f_mag}
\end{figure}

The target region in the photosphere is dominated by the negative polarity, and the positive polarity is scattered around in small patches. 
Another small positive polarity flux emerges before the jet time, which is likely to be associated with the jetting. 
We attempt to infer the overall magnetic connectivity from the \ha\ spicule trajectory assuming they follow the magnetic field to some extent.  
%While X-ray or EUV images can be used too, the EUV jetting  structure in the present case arises so weak, limited, and spatially confined area.
In Figure \ref{f_mag}$a$, we plot 
the GST/VIS \ha\ blue wing image so that the spicules appear as the white narrow features and the HMI line-of-sight (LOS) magnetic field as contours. The green curves are readouts of the spicule trajectories from the \ha\ blue wing image, which are then copied to the other panels: the AIA 171 \AA\ intensity image (Figure \ref{f_mag}$b$), and the HMI LOS magnetogram itself (Figure \ref{f_mag}$d$), and as magenta lies over the Dopplergram (Figure \ref{f_mag}$c$).
The spicule trajectories (green curves) outline the magnetic field stemming from the negative polarity flux running toward the positive polarity patch far off in the north. 
%Since spicules are numerous and constantly replenished the set of field lines implies an arcade-like magnetic structure that connects the main negative polarity patches to the northern positive patches.

An important implication of the \ha\ spicule trajectories would be that they map the magnetic field lines connecting the EUV jet base to the photosphere and that of the EUV jet spire to remote regions. With the AIA image alone, we do not see the EUV jet reach down to the emerging positive flux patch. However, the \hri\ images show the EUV jet being extended further south, and the \ha\ spicule trajectories indicate the magnetic field lines rapidly wrapped toward the positive flux patch underneath. Furthermore, the red-shifted \ha\ components are densely packed under the EUV jets as if they are \ha\ counterpart of JBPs.
The upward trajectory of the EUV jet also follows this curved path of the {\ha} spicules as shown in Figure \ref{f_mag}$b$.
Of course, not all field lines are exactly parallel to each other with inclination angles varying with position to some extent, but the overall EUV jet trajectories tend to follow the westward structure implied by the \ha\ spicules. The westward wrapping structure is also obvious in the Dopplergram (Figure \ref{f_mag}$c$). 
%Whereas the spicules (taken from the blue wing images) tend to align with the blue-shifted features, the strong red-wing features are concentrated in one footpoint of the inferred loop rather than exactly located at the PIL.  
The field lines stemming from the positive flux patch seem to be the outer most of this overarching magnetic structure.

The HMI magnetogram in Figure \ref{f_mag}$d$ shows a few  positive polarity patches in the north. They are probably connected to the the negative flux region to form an arcade-like structure, as implied by the spicule trajectories. A large patch in the positive polarity exists in the east, but these field lines are connected to the east in view of the \ha\ images. Beneath the EUV jets, the photospheric magnetic fields were of the negative polarity alone, but at 19:06 UT, a small positive flux patch emerges in the south (within the cyan box), and the EUV jets started. 

Figure \ref{f_mag}$e$ shows NIRIS magnetogram in the small FOV indicated by the cyan box in Figure \ref{f_mag}$d$ in order to investigate the evolution of the positive polarity patch. This small structure is not clearly visible in the HMI magnetogram, because of the limited resolution and sensitivity. With the emergence of the small positive flux patch at \sm19:07 UT, it becomes a mixed polarity region. The positive-polarity flux kept increasing without an obvious signature of magnetic flux cancellation. When it converged toward the PIL and the jets have occurred.
In Figure \ref{f_mag}$f$, we set two slits (solid and dashed) to read out one-dimensional (1-D) magnetic profiles shown in Figure \ref{f_mag}$g$.
The field strength corresponds to the local vertical magnetic field, $B_z$, obtained by dividing the LOS magnetic field by $\cos\theta$ with the position angle, $\theta=27^\circ$ from the disk center. The separation between a nearby bipolar pair is estimated as 1.6 Mm, and their peak strengths are $+$80 G and $-$180 G, respectively, indicating that it is one of typical smallest SSEs \citep{Parnell+etal2009ApJ...698...75P, Lee2025ApJ...988L..16L}.
%The evolving complicated EUV intensity stripe pattern should be associated with a bundle of newly reconnected field lines stemming from bipolar patches in the negative polarity. Whether this represents either instantly forming flux ropes or {\alf} wave train is an important issue to investigate.

\subsection{Plasma Properties and Energies of the EUV Jets}\label{dem}

% revised
The temperature and electron density diagnostics for the EUV features are usually available from the differential emission measure (DEM) analysis of SDO/AIA \citep{hannah2012differential, cheung2015thermal}. However, not only is the AIA EUV intensity of the present target low, but there is also a bright 171 \AA\ EUV loop located in the foreground of the jet, which may add unwanted contamination to the jet source. For this reason, we compare the DEMs derived from EUI/Full Sun Imager (FSI) images, which are less contaminated than features from AIA. \lee{FSI provides only two channels, which are practically inadequate for constraining DEM solutions. 
%It is desirable to predict 6 channel AIA-equivalent images from the two channel FSI images. 
\citet{youn2025can} utilized a deep learning model based on Pix2PixCC \citep{jeong2022improved} to be able to reproduce the five AIA channel (94, 193, 211, 131, 335\AA) images from FSI 174 and 304 \AA\ images. 
They determined the DEM by applying the standard procedure to this full set of the five AIA-equivalent images plus the original FSI 174 \AA\ images. The resulting DEMs agrees well with those derived from actual AIA observations across various temperature ranges. We therefore applied this technique to the FSI data to obtain the DEM results presented here.
}

\begin{figure}[ht]
\centering
\includegraphics[width=\textwidth]{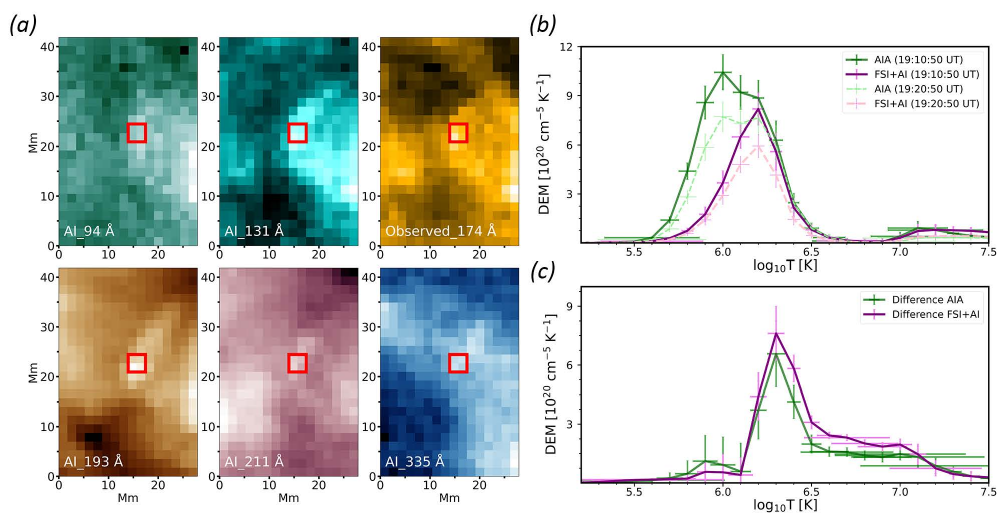}
\caption  {DEMs of the jet region derived from EUV images.
\lee{(a)} The six panels show one EUI/FSI 174 {\AA} image and five AIA equivalent images generated by deep learning. The red box indicates the pixels used to calculate the DEM. \lee{(b)} The DEM derived from the SDO/AIA images (green) with AI-generated EUV images with observed FSI 174 {\AA} image (purple). The solid lines represent the DEM profiles when the jet activity is strong (19:10:20 UT), while the dashed lines correspond to the profiles when the jet diminished (19:20:50 UT). \lee{(c) The net DEM calculated from the difference EUV images between the jet maximum and the jet quiet time.}}
\label{jmyoun}
\end{figure}

% before revise
% The temperature and electron density diagnostics for the EUV features are available from the differential emission measure (DEM) analysis of SDO/AIA \citep{hannah2012differential, cheung2015thermal}. However, not only that the AIA EUV intensity in the present target is low, but that a bright 171 \AA\ EUV loop is located in the foreground of the jet, it may possibly add unwanted contamination to the jet source.
% %The foreground loop is well visible in the AIA 171, 193, 131 \AA\ channels and less in the 94, 211, 335 {\AA} channels.
% For this reason, we utilize the EUI/Full Sun Imager (FSI) images. They prove only two channels, and do not allow DEM analysis. However, a recently developed tool can reproduce AIA equivalent images by training deep learning models based on Pix2PixCC \citep{youn2025can}. 
Figure \ref{jmyoun} shows the AIA-like five-channel dataset generated using deep learning (hereafter AI-EUI), the EUV images observed by EUI$_{\rm FSI}$ at 174 \AA\, and the corresponding derived DEMs. The error bars from the DEM plots are given by \cite{hannah2012differential} method, using AIA and EUI data manuals \citep{boerner2012initial, Rochus_2020}, \lee{and include the errors from the jet maximum period (19:10 UT) images and those of the jet quiet time (19:20 UT) images as well.} At the jet maximum period (19:10 UT), the AIA DEM shows a peak at $\log T \approx 6.0$, whereas the AI-EUI DEM shows a peak at $\log T \approx 6.2$. The former peak includes an extra component coming from the foreground coronal loop in the AIA's viewpoint, whereas the latter must be the temperature of the plasma surrounding the jet. There is another peak of DEM at $\log T \approx 7.1$. This high-temperature peak is found for both AIA and AI-EUI and is obvious at 19:10 UT when the jet activity is strongest. The DEM around this temperature becomes low when the jets diminished at 19:20 UT, evidencing that this must be the temperature range of the hot jet.

We subtract the EUI/FSI intensity at a jet quiet time (19:20:50 UT) from that at the jet time (19:10:50 UT) to obtain the net DEM of the jet.
Since the jets under current study are small and faint feature, the background subtraction is tricky \citep[cf.][]{Zhang2014}.
We thus keep comparing AI-EUI results with those of AIA to check if the same trend is found in both results. If they behave similarly, we finally adopt the AI-EUI's results for DEM. 
We then calculate, from the net DEM, the emission measure of the jets, $EM = \sum_{i}DEM(\log T_i) \Delta \log T_i$ in the temperature range $4.0\leq \log T \leq 8.0$ in the interval of $\Delta \log T = 0.1$. The median EM value over the jet area is calculated as $5.5\times10^{27}$ cm$^{-5}$ (lower right panel of Figure \ref{jmyoun}). 

Note that the jet is so small and its area on the EUI/FSI image is only four pixels wide. To determine the geometrical factors of the jet, we can no longer use EUI/FSI images, but the {\hri} images instead. Although the jet is of highly entangled structure, we find 30\% contour level of the maximum intensity of the \hri\ 174 {\AA} image well outlines the jet area. On the {\hri} image, we count 475 pixels within the jet area, which amounts to $A\approx 1.0\times 10^{17}$ cm$^2$. The average width of the jet strands is about 5 pixels or $w\approx 7.3\times 10^7$ cm. In the regions where individual strands are resolved we take $w$ as the same as their LOS thickness, $h$. In the lower part of the jets, many strands are entangled, where we increase $h$ according to the area of the brightened area. 
In this way we determine total number of electrons, $N=(EM/h)^{1/2}Ah\approx 2.0\times 10^{35}$ and total mass of the fully ionized plasma, $M=1.1Nm_H\approx 3.7\times 10^{11}$ g.
%$A$ as $h=A^{1/2}$ \citep{joulin2016energetic} \textcolor{red}{$h\approx 1744$ km}.
Thermal energy of the jet is then estimated as
%$E_{th} =  3k_BTA(EMfh)^{1/2}$, 
$3Nk_B\langle T\rangle\approx 1.7\times 10^{26}$ erg, where $k_B$ is the Boltzmann constant and $\langle T \rangle$ is the DEM-weighted temperature. Kinetic energy of the jet is $Mv^2/2\approx 2.2 \times 10^{25}$ erg, using the mass, $M$, and the jet speed, $v=110$ km s$^{-1}$. Total energy of the EUV jet therefore amounts to $1.9 \times 10^{26}$ erg with 87\% in the thermal energy and 13\% in the kinetic energy.

\section{Discussion}\label{model}

This rare collaboration between EUI/HRI and GST/VIS presents a picture that the upwardly propagating EUV jets in the corona and the \ha\ redshift in the chromosphere are adjoined at the the X-point. The hypothesis of the interchange reconnection in the corona is supported by the location and timing of multipolarity patches in the photosphere without magnetic flux cancellation and footpoint brightenings. 
However, straight trajectories of the ambient \ha\ spicules are in sharp contrast with the helical EUV jets, posing a puzzle regarding the role of magnetic reconnection in coronal jets and chromospheric ejections.
%There is, at least, a general trend of {\ha} being oriented wrapped around and run toward the west. The trajectory of the EUV jets is following this path of the {\ha} spicules, and thus representing the overall magnetic structure there. 
We  briefly discuss this issue based on two highly relevant models for small-scale jets, both of which self-consistently incorporate the chromosphere and corona within a unified framework  \citep{Iijima2017, Longcope2025}.

The chromospheric jet model proposed by \citet{Iijima2017} includes the photospheric radiative transfer and the equations of state in the simulation of the chromospheric jets.
The most attractive feature of this model is that it predicts formation jets in the scales close to the present observation, specifically, the simulated jets consist of finer strands packed with a maximum height of 10–11 Mm and lifetime of 8–10 minutes in a cluster with a diameter of several Mm. The jets in this model are driven by the Lorentz force from the magnetic field lines strongly entangled in the chromosphere, which is somewhat inconsistent with the straight trajectories of \ha\ spicules although more appropriate for the EUV jet structure in the corona.

\begin{figure}[ht]
    \centering
    \includegraphics[width=\textwidth]{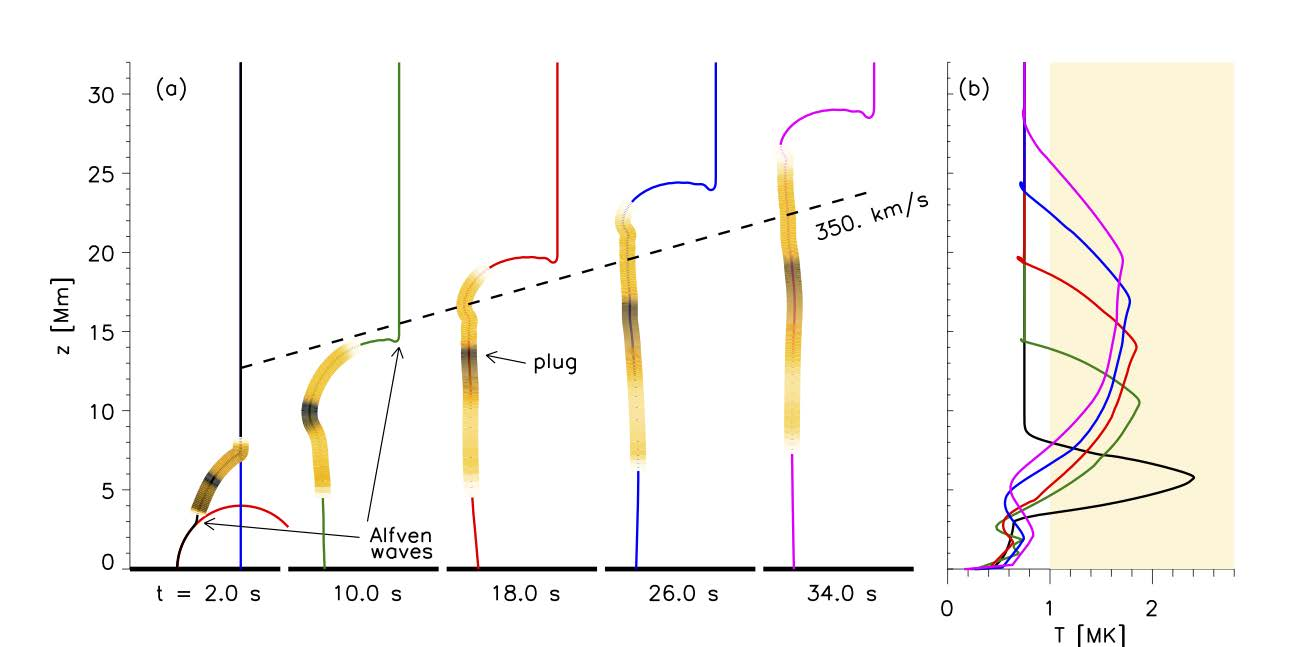}\\
    \includegraphics[width=\textwidth]{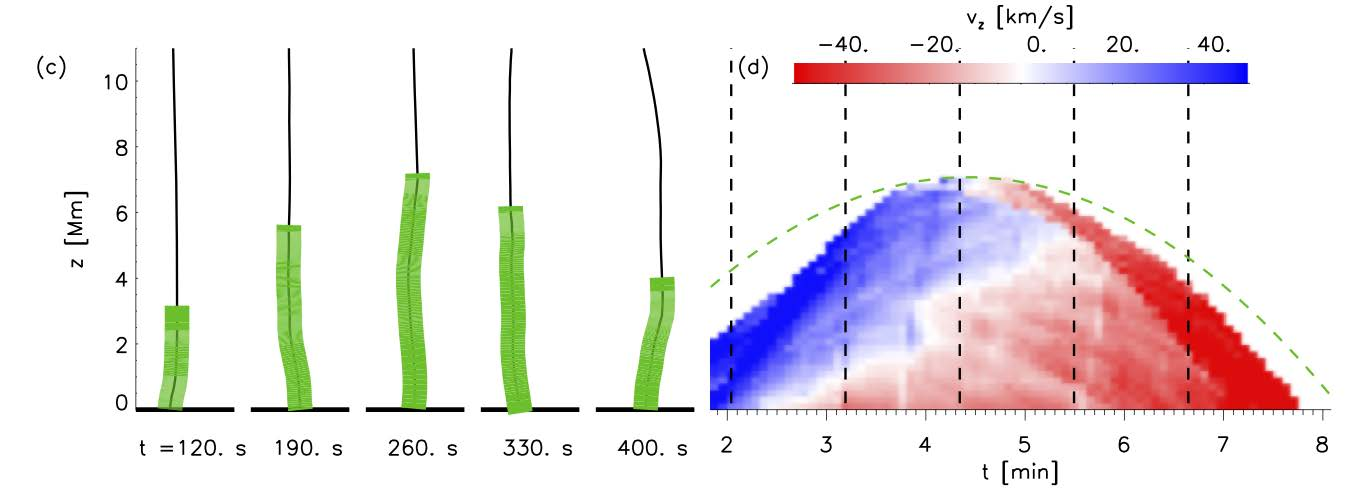}
\caption{The reconnection model illustrated using a thin flux tube simulation.  (a) Five snapshots from the initial phase in which the field line (colored curve) relaxes following its formation by reconnection.  The pre-reconnection field lines are shown as red and blue curves on the left-most snapshot.  Reverse color scale shows emission measure of hot ($T>10^6$ K) plasma.  A dashed diagonal line shows a constant velocity, for reference. 
 (b) The temperature {\em vs.}\ height for the five times using the same colors.  The shaded box indicates the hot plasma whose emission is shown in (a).   
 %(c) Shows five later times at which 
 (c) At five later times, a spicule has been launched by the downward Alfv\'en wave.  The green region shows cool ($<10^5$ K), dense plasma at each time.  
 % (d) Shows the 
(d) The vertical velocity (shaded) in space-time coordinates.  A green dashed parabola shows free-fall for reference, and vertical dashed lines show the times of the snapshots.} 
\label{a1}
\end{figure}

The reconnection model of \citet{Longcope2025} considers impulsive, localized interchange reconnection in the low corona, as illustrated in Figure \ref{a1}.  Following its formation by reconnection, the field line straightens under tension (i.e.\ the Lorentz force).  The straightening process consists of large-amplitude Alfv\'en waves propagating up and down, away from the reconnection site (see Figure \ref{a1}a).  The propagating waves produce parallel flows which collide to form a high-temperature, high density, central plug between them \citep{Longcope2009,Longcope2011,Longcope2015}.  The reconnection between a closed and open field line (red semi-circle and blue vertical line respectively, on the left snapshot of Figure \ref{a1}a) is asymmetric due to the significant plasma pressure difference between them.  This asymmetry causes the plug to move upward, although at a speed somewhat slower than the Alfv\'en speed in either field.  (Note how the upper bend in Figure \ref{a1}a, runs past the dashed line.)

The downward {\alf} wave impacts the chromosphere where it drives cool, dense plasma upward to form a spicule \citep{Longcope2025} -- see Figure \ref{a1}c.  This occurs later, when the field line has almost entirely relaxed, leading to a much straighter plasma than the EUV loop.  The upflow is much slower (by almost an order of magnitude in the illustrative example), causing it to reach its apex 4 mins.\ after reconnection.

The model is illustrated in Figure \ref{a1} using a thin flux tube simulation \citep{Guidoni2010,Longcope2015,Longcope2025} whose parameters have not been tuned to match the observation.  Nevertheless, the example shows the rough magnitudes of its various components.  The hot, dense plug moves upward at a fraction of the Alfv\'en speed, consistent with the 110 km/s that the observed EUV jets move across the plane of the sky.  Moreover, its structuring by the evolving post-reconnection field gives the EUV loop an angle relative to the dominant, open field.  Our simulation considers evolution of untwisted field lines confined to a plane \citep{Longcope2025}, leaving the EUV jet with a distortion, but not helical twist.  A more complicated scenario, with twist in the initial and/or final field, could exhibit more properly helical distortion in the EUV loops during this initial phase of relaxation.

The same simulation exhibits a spicule when cool, dense plasma is launched upward by the impact of the downward {\alf} wave on the chromosphere.  The cool plasma travels at $\sim40$ km/s, reaching a peak of 7 Mm before falling back under gravity.  This is in line with the Doppler shifts and lengths of the {\ha} observations. The cooler plasma follows the relaxed field, making it much straighter,and more vertical, than the EUV loop, consistent with observation.

Under this model, reconnection releases energy by decreasing the length of the field line by $\Delta L$.  A closed loop with mean field strength $B_0$, reconnecting to open field, will release $W_{\rm mag}/\Phi=B_0\Delta L/4\pi$ \citep{Longcope2015,Longcope2025}.  For a rough estimate we consider a loop connecting the parasitic polarity to the nearest negative region
%, 2.2 Mm away.  This separation maps to a semicircle of radius $r=1.1$ Mm, which would therefore be the height of the reconnection point.  Due to this very low loop, we take $B_0\simeq130$ G, the average between the photospheric field strengths of the two polarity peaks ($\sim +80$ G and $-180$ G).  Transforming a quarter circle to a vertical segment of the same height, $r$, shortens the tube by $\Delta L=(\pi/2-1)r=0.63$ Mm.  Such a shortening releases $W_{\rm mag}/\Phi\simeq 6.5\times 10^8\, {\rm erg/Mx}$.  The $w=0.73$ Mm diameter loop will have flux $\Phi=B_0\pi(w/2)^2=5.4\times 10^{17}$ Mx, and therefore receive $W_{\rm mag}=3.5\times 10^{26}\, {\rm erg}$ from reconnection.  
with a measured separation of 1.6 Mm (Figure \ref{f_mag}$g$).
This separation maps to a semicircle of radius $r=0.80$ Mm, which would therefore be the height of the reconnection point.  Due to this very low loop, we take $B_0\simeq130$ G, the average between the photospheric field strengths of the two polarity peaks ($\sim +80$ G and $-180$ G).  Transforming a quarter circle to a vertical segment of the same height, $r$, shortens the tube by $\Delta L=(\pi/2-1)r=0.46$ Mm.  Such a shortening releases $W_{\rm mag}/\Phi\simeq 5.3\times 10^8\, {\rm erg/Mx}$.  The $w=0.73$ Mm diameter loop will have flux $\Phi=B_0\pi(w/2)^2=5.4\times 10^{17}$ Mx, and therefore receive $W_{\rm mag}=2.6\times 10^{26}\, {\rm erg}$ from reconnection.  
This will be shared between the EUV jet and the spicules, and is therefore roughly  consistent with our estimate of the energy in the jet alone (Section \ref{dem}).

\section{Concluding Remarks}

The first joint observations of small-scale EUV jets using the SolO's \hri\ and BBSO's GST/VIS \ha\ during the October 2022 Campaign for PSP, we had a unique opportunity to directly compare the small, short-duration EUV jets with abundant \ha\ spicules around in unprecedented details.  
The observations and diagnostic results are summarized in the abstract and will not be repeated here.
Here we wish to remark on the main issue: what dictates the structure and dynamics of the resulting SSEs in the chromosphere and the corona. The interchange reconnection as a trigger of the jet activity is unambiguously determined from the  location of the EUV jets and \ha\ spicules relative to the photospheric magnetic fields,
However, 
we encountered two unexpected puzzles:
how the entangled structure forms and why it exclusively appears in the EUV wavelengths not in the {\ha} line.
All \ha\ features accompanying the EUV jets including the sheath flow, bulk downflow, and normal upward spicules exhibit straight trajectories.

Among the alternatives, the reconnection model, which begins with interchange reconnection, suggests some explanations.  The twisted, or non-aligned, EUV structures form promptly in the corona as a result of the reconnection energy release there.  Prompt heating and compressions makes visible plasma which had previously not been hot enough or dense enough to appear in EUV.  Our demonstration, with a simplified geometry, exhibits a distorted EUV loop, but a more complicated configuration could exhibit helices.  The compressed and heated plasma plug will move along the reconnected loop at a speed comparable to, but somewhat slower than, the Alfv\'en speed, in agreement with our observation.  Finally, the cooler \ha\ feature forms through a secondary response to the episode of interchange reconnection, after the downward \alf\ wave impacts the chromosphere.  This plasma moves upward much more slowly, and along the field that has already fully relaxed to become straight and vertical. The plasma at the feet appears to be red-shifted at first, owing to chromospheric condensation, and later when the cool plasma drains back down. 
%This interchange reconnection occurs in a loop in a loop with magnetic flux \sm$10^{17}$ Mx and separation \sm2$''$, belonging to the smallest magnetic elements detectable in the quiet Sun \citep{Parnell+etal2009ApJ...698...75P} and produces energy \sm$10^{26}$ erg corresponding to typical smallest SSEs \citep{Lee2025ApJ...988L..16L}.

\lee{To compare with other small jets detected by \hri,
the picoflare jets have kinetic energies below $10^{24}$ erg \citep{Chitta2023Sci...381..867C}, at least, two orders of magnitude lower than that of the jet studied here. Despite this energy discrepancy, they share similar spatial scales, lifetimes, and  velocities.
Similarly, the helical \hri\ jet studied by \citet{Petrova2024A&A...687A..13P} displays a comparable morphology and was also interpreted as propagating torsional \alf\ waves in a low coronal structure with estimated wave energy flux of 140 kW m$^{-2}$. If we tentatively calculate the corresponding kinetic energy of the {\alf}ic pulse with assumed loop diameter of 2 Mm and lifetime of 2 min, we find $5.3\times 10^{26}$ erg.
In contrast, we estimated the total energy of the present jet as \sm$1.9 \times 10^{26}$ erg with 87\% in thermal energy and 13\% in kinetic energy based on the DEM analysis as well as the NIRIS measurements of the footpoint separation \sm2$''$ and magnetic flux \sm$10^{17}$ Mx. All these jets are interpreted as powered by magnetic reconnection at the smallest scales of magnetic elements detectable in the quiet Sun \citep{Parnell+etal2009ApJ...698...75P, Lee2025ApJ...988L..16L}.
}

% If the bright strands in the EUV jet, with helical structure, is a manifestation of a magnetic flux rope, we have to explain how it can instantly form and disappear, and why the \ha\ plasma does not move along the twisted magnetic structure if they exist? The only \ha\ counterpart of the EUV jets are the ripples around the jet surface. Such instantaneous formation of the helical structure and its short duration are no issue if the entanglement is a manifestation of {\alf}ic wave front. The energy associated with the jets predicted by the {\alf} model is comparable to the energies derived from DEM analysis to demonstrate the ability of the mechanism for the jets. It is also important that the model predicts that interchange reconnection can transfer energy through such a narrow channel as narrow as the bipolar separation underneath to explain how the EUV jets could be so isolated in space and time amid ubiquitous spicules.  Eventually such differences between \ha\ spicules and the EUV jets in the spatio-temporal distribution along with the morphology  raises a question regarding why they look so different from each other if commonly driven by magnetic reconnection. It is possible that the spicules may not originate from magnetic reconnection at all, 
%or, at least, the same type of reconnection. 
% challenging the conventional idea.
 
The results of this collaboration highlights the critical role of high-resolution GST observations in bridging the gap between the photosphere and corona by providing spectroscopic data on chromospheric dynamics alongside fine-structured coronal jets. These findings boosts SolO's objective of advancing our understanding of how the Sun generates small-scale ejections in the chromosphere and corona, ultimately enhancing models of solar-heliospheric connections.

\begin{acknowledgments}
We wish to thank the referee for very thoughtful and helpful comments. We gratefully acknowledge the use of data from the Goode Solar Telescope (GST) of the Big Bear Solar Observatory (BBSO). BBSO operation is supported by US NSF AGS-2309939 grant and New Jersey Institute of Technology. GST operation is partly supported by the Korea Astronomy and Space Science Institute and the Seoul National University.
\lee{Solar Orbiter is a space mission of international collaboration between ESA and NASA, operated by ESA. The EUI instrument was built by CSL, IAS, MPS, MSSL/UCL, PMOD/WRC, ROB, LCF/IO with funding from the Belgian Federal Science Policy Office (BELSPO/PRODEX PEA 4000112292); the Centre National d’Etudes Spatiales (CNES); the UK Space Agency (UKSA); the Bundesministerium für Wirtschaft und Energie (BMWi) through the Deutsches Zentrum für Luft- und Raumfahrt (DLR); and the Swiss Space Office (SSO).} 
This work was supported by NSF grants, AGS-2114201, AGS-2229064 and AGS-2309939, and NASA grants, 80NSSC19K0257, 80NSSC20K0025, 80NSSC20K1282, 80NSSC24K1914 and 80NSSC24K0258. NKP acknowledges support from NASA's SDO/AIA grant and NSF AAG award (no. 2307505). We acknowledge the use of  SDO/AIA/HMI data. AIA is an instrument onboard the Solar Dynamics Observatory, a mission for NASA’s Living With a Star program.
\end{acknowledgments}

\vspace{5mm}
\facilities{SO (EUI), SDO (AIA, HMI), BBSO (GST/VIS and GST/NIRIS)}

\software{IDL, SolarSoft \citep{2012ascl.soft08013F},
%, SWAMIS \citep{DeForest+etal2007ApJ...666..576D}, 
OF and DopplerCG \citep{2022XYang}}


\begin{thebibliography}{}
\expandafter\ifx\csname natexlab\endcsname\relax\def\natexlab#1{#1}\fi
\providecommand{\url}[1]{\href{#1}{#1}}
\providecommand{\dodoi}[1]{doi:~\href{http://doi.org/#1}{\nolinkurl{#1}}}
\providecommand{\doeprint}[1]{\href{http://ascl.net/#1}{\nolinkurl{http://ascl.net/#1}}}
\providecommand{\doarXiv}[1]{\href{https://arxiv.org/abs/#1}{\nolinkurl{https://arxiv.org/abs/#1}}}

\bibitem[{{Beckers}(1977)}]{beckers77}
{Beckers}, J.~M. 1977, in Astrophysics and Space Science Library, Vol.~69, Illustrated Glossary for Solar and Solar-Terrestrial Physics, ed. A.~{Bruzek} \& C.~J. {Durrant}, 21, \dodoi{10.1007/978-94-010-1245-4_4}

\bibitem[{{Berghmans} {et~al.}(2021){Berghmans}, {Auch{\`e}re}, {Long}, {Soubri{\'e}}, {Mierla}, {Zhukov}, {Sch{\"u}hle}, {Antolin}, {Harra}, {Parenti}, {Podladchikova}, {Aznar Cuadrado}, {Buchlin}, {Dolla}, {Verbeeck}, {Gissot}, {Teriaca}, {Haberreiter}, {Katsiyannis}, {Rodriguez}, {Kraaikamp}, {Smith}, {Stegen}, {Rochus}, {Halain}, {Jacques}, {Thompson}, \& {Inhester}}]{Berghmans_2021}
{Berghmans}, D., {Auch{\`e}re}, F., {Long}, D.~M., {et~al.} 2021, \aap, 656, L4, \dodoi{10.1051/0004-6361/202140380}

\bibitem[{{Bizien} {et~al.}(2025){Bizien}, {Froment}, {Madjarska}, {Dudok de Wit}, \& {Velli}}]{Bizien2025A&A...694A.181B}
{Bizien}, N., {Froment}, C., {Madjarska}, M.~S., {Dudok de Wit}, T., \& {Velli}, M. 2025, \aap, 694, A181, \dodoi{10.1051/0004-6361/202452140}

\bibitem[{Boerner {et~al.}(2012)Boerner, Edwards, Lemen, Rausch, Schrijver, Shine, Shing, Stern, Tarbell, Title, {et~al.}}]{boerner2012initial}
Boerner, P., Edwards, C., Lemen, J., {et~al.} 2012, The Solar Dynamics Observatory, 41

\bibitem[{{Bohlin} {et~al.}(1975){Bohlin}, {Vogel}, {Purcell}, {Sheeley}, {Tousey}, \& {Vanhoosier}}]{Bohlin1975_macrospics}
{Bohlin}, J.~D., {Vogel}, S.~N., {Purcell}, J.~D., {et~al.} 1975, \apjl, 197, L133, \dodoi{10.1086/181794}

\bibitem[{{Brueckner}(1982)}]{Brueckner1982}
{Brueckner}, G.~E. 1982, in Bulletin of the American Astronomical Society, Vol.~14, 411--415

\bibitem[{{Cheung} {et~al.}(2015){Cheung}, {Boerner}, {Schrijver}, {Testa}, {Chen}, {Peter}, \& {Malanushenko}}]{cheung2015thermal}
{Cheung}, M. C.~M., {Boerner}, P., {Schrijver}, C.~J., {et~al.} 2015, \apj, 807, 143, \dodoi{10.1088/0004-637X/807/2/143}

\bibitem[{{Chitta} {et~al.}(2023){Chitta}, {Zhukov}, {Berghmans}, {Peter}, {Parenti}, {Mandal}, {Aznar Cuadrado}, {Sch{\"u}hle}, {Teriaca}, {Auch{\`e}re}, {Barczynski}, {Buchlin}, {Harra}, {Kraaikamp}, {Long}, {Rodriguez}, {Schwanitz}, {Smith}, {Verbeeck}, \& {Seaton}}]{Chitta2023Sci...381..867C}
{Chitta}, L.~P., {Zhukov}, A.~N., {Berghmans}, D., {et~al.} 2023, Science, 381, 867, \dodoi{10.1126/science.ade5801}

\bibitem[{{Cook} {et~al.}(1983){Cook}, {Brueckner}, \& {Bartoe}}]{Cook1983}
{Cook}, J.~W., {Brueckner}, G.~E., \& {Bartoe}, J. D.~F. 1983, \apjl, 270, L89, \dodoi{10.1086/184076}

\bibitem[{{De Pontieu} {et~al.}(2014){De Pontieu}, {Title}, {Lemen}, {et~al.}}]{DePontieu.IRIS.2014SoPh..289.2733D}
{De Pontieu}, B., {Title}, A.~M., {Lemen}, J.~R., {et~al.} 2014, \solphys, 289, 2733, \dodoi{10.1007/s11207-014-0485-y}

\bibitem[{{Dere} {et~al.}(1989{\natexlab{a}}){Dere}, {Bartoe}, \& {Brueckner}}]{Dere1989}
{Dere}, K.~P., {Bartoe}, J. D.~F., \& {Brueckner}, G.~E. 1989{\natexlab{a}}, \solphys, 123, 41, \dodoi{10.1007/BF00150011}

\bibitem[{{Dere} {et~al.}(1989{\natexlab{b}}){Dere}, {Bartoe}, {Brueckner}, {Cook}, {Socker}, \& {Ewing}}]{Dere1989_macrospic}
{Dere}, K.~P., {Bartoe}, J. D.~F., {Brueckner}, G.~E., {et~al.} 1989{\natexlab{b}}, \solphys, 119, 55, \dodoi{10.1007/BF00146212}

\bibitem[{{Freeland} \& {Handy}(2012)}]{2012ascl.soft08013F}
{Freeland}, S.~L., \& {Handy}, B.~N. 2012, {SolarSoft: Programming and data analysis environment for solar physics}, Astrophysics Source Code Library, record ascl:1208.013

\bibitem[{{Goode} {et~al.}(2010){Goode}, {Yurchyshyn}, {Cao}, {Abramenko}, {Andic}, {Ahn}, \& {Chae}}]{goode10}
{Goode}, P.~R., {Yurchyshyn}, V., {Cao}, W., {et~al.} 2010, \apjl, 714, L31, \dodoi{10.1088/2041-8205/714/1/L31}

\bibitem[{{Guidoni} \& {Longcope}(2010)}]{Guidoni2010}
{Guidoni}, S.~E., \& {Longcope}, D.~W. 2010, \apj, 718, 1476, \dodoi{10.1088/0004-637X/718/2/1476}

\bibitem[{{Hannah} \& {Kontar}(2012)}]{hannah2012differential}
{Hannah}, I.~G., \& {Kontar}, E.~P. 2012, \aap, 539, A146, \dodoi{10.1051/0004-6361/201117576}

\bibitem[{{Iijima} \& {Yokoyama}(2017)}]{Iijima2017}
{Iijima}, H., \& {Yokoyama}, T. 2017, \apj, 848, 38, \dodoi{10.3847/1538-4357/aa8ad1}

\bibitem[{{Jeong} {et~al.}(2022){Jeong}, {Moon}, {Park}, {Lee}, \& {Baek}}]{jeong2022improved}
{Jeong}, H.-J., {Moon}, Y.-J., {Park}, E., {Lee}, H., \& {Baek}, J.-H. 2022, \apjs, 262, 50, \dodoi{10.3847/1538-4365/ac8d66}

\bibitem[{{Karovska} \& {Habbal}(1994)}]{Karovska1994_macrospic}
{Karovska}, M., \& {Habbal}, S.~R. 1994, \apjl, 431, L59, \dodoi{10.1086/187472}

\bibitem[{{Lee} {et~al.}(2025){Lee}, {Georgoulis}, {Sharma}, {Raouafi}, {Li}, \& {Wang}}]{Lee2025ApJ...988L..16L}
{Lee}, J., {Georgoulis}, M.~K., {Sharma}, R., {et~al.} 2025, \apjl, 988, L16, \dodoi{10.3847/2041-8213/adeb54}

\bibitem[{{Lee} {et~al.}(2022){Lee}, {Yurchyshyn}, {Wang}, {Yang}, {Cao}, \& {Martínez Oliveros}}]{Lee_2022}
{Lee}, J., {Yurchyshyn}, V., {Wang}, H., {et~al.} 2022, \apj, 935L, 27

\bibitem[{Longcope \& Klaassen(2025)}]{Longcope2025}
Longcope, D., \& Klaassen, P. 2025, 989, 152, \dodoi{10.3847/1538-4357/adf0ef}

\bibitem[{{Longcope} \& {Guidoni}(2011)}]{Longcope2011}
{Longcope}, D.~W., \& {Guidoni}, S.~E. 2011, 740, 73, \dodoi{10.1088/0004-637X/740/2/73}

\bibitem[{Longcope {et~al.}(2009)Longcope, Guidoni, \& Linton}]{Longcope2009}
Longcope, D.~W., Guidoni, S.~E., \& Linton, M.~G. 2009, 690, L18, \dodoi{10.1088/0004-637X/690/1/L18}

\bibitem[{{Longcope} \& {Klimchuk}(2015)}]{Longcope2015}
{Longcope}, D.~W., \& {Klimchuk}, J.~A. 2015, \apj, 813, 131, \dodoi{10.1088/0004-637X/813/2/131}

\bibitem[{{Mandal} {et~al.}(2023){Mandal}, {Peter}, {Chitta}, {Cuadrado}, {Sch{\"u}hle}, {Teriaca}, {Solanki}, {Harra}, {Berghmans}, {Auch{\`e}re}, {Parenti}, {Zhukov}, {Buchlin}, {Verbeeck}, {Kraaikamp}, {Rodriguez}, {Long}, {Schwanitz}, {Barczynski}, {Pelouze}, {Smith}, {Liu}, \& {Cheung}}]{Mandal2023}
{Mandal}, S., {Peter}, H., {Chitta}, L.~P., {et~al.} 2023, \aap, 670, L3, \dodoi{10.1051/0004-6361/202245431}

\bibitem[{{Moore} {et~al.}(1977){Moore}, {Tang}, {Bohlin}, \& {Golub}}]{Moore1977}
{Moore}, R.~L., {Tang}, F., {Bohlin}, J.~D., \& {Golub}, L. 1977, \apj, 218, 286, \dodoi{10.1086/155681}

\bibitem[{{Müller} {et~al.}(2020){Müller}, {St. Cyr}, {Zouganelis}, {et~al.}}]{Muller_2020}
{Müller}, D., {St. Cyr}, O.~C., {Zouganelis}, I., {et~al.} 2020, \aap, 642, A1

\bibitem[{{Panesar} {et~al.}(2023){Panesar}, {Hansteen}, {Tiwari}, {Cheung}, {Berghmans}, \& {Müller}}]{Panesar_2023}
{Panesar}, N.~K., {Hansteen}, V. H.~H., {Tiwari}, S.~K., {et~al.} 2023, \apj, 943, 24

\bibitem[{{Panesar} {et~al.}(2018){Panesar}, {Sterling}, {Moore}, {Tiwari}, {De Pontieu}, \& {Norton}}]{Panesar2018}
{Panesar}, N.~K., {Sterling}, A.~C., {Moore}, R.~L., {et~al.} 2018, \apjl, 868, L27, \dodoi{10.3847/2041-8213/aaef37}

\bibitem[{{Panesar} {et~al.}(2021){Panesar}, {Tiwari}, {Berghmans}, {Cheung}, {Müller}, {Auchere}, \& {Zhukov}}]{Panesar_2021}
{Panesar}, N.~K., {Tiwari}, S.~K., {Berghmans}, D., {et~al.} 2021, \apj, 921L, 20

\bibitem[{{Panesar} {et~al.}(2019){Panesar}, {Sterling}, {Moore}, {Winebarger}, {Tiwari}, {Savage}, {Golub}, {Rachmeler}, {Kobayashi}, {Brooks}, {Cirtain}, {De Pontieu}, {McKenzie}, {Morton}, {Peter}, {Testa}, {Walsh}, \& {Warren}}]{Panesar2019}
{Panesar}, N.~K., {Sterling}, A.~C., {Moore}, R.~L., {et~al.} 2019, \apjl, 887, L8, \dodoi{10.3847/2041-8213/ab594a}

\bibitem[{{Parnell} {et~al.}(2009){Parnell}, {DeForest}, {Hagenaar}, {Johnston}, {Lamb}, \& {Welsch}}]{Parnell+etal2009ApJ...698...75P}
{Parnell}, C.~E., {DeForest}, C.~E., {Hagenaar}, H.~J., {et~al.} 2009, \apj, 698, 75, \dodoi{10.1088/0004-637X/698/1/75}

\bibitem[{{Petrova} {et~al.}(2024){Petrova}, {Van Doorsselaere}, {Berghmans}, {Parenti}, {Valori}, \& {Plowman}}]{Petrova2024A&A...687A..13P}
{Petrova}, E., {Van Doorsselaere}, T., {Berghmans}, D., {et~al.} 2024, \aap, 687, A13, \dodoi{10.1051/0004-6361/202348799}

\bibitem[{{Raouafi} \& {Stenborg}(2014)}]{raouafi14}
{Raouafi}, N.-E., \& {Stenborg}, G. 2014, \apj, 787, 118, \dodoi{10.1088/0004-637X/787/2/118}

\bibitem[{{Raouafi} {et~al.}(2016){Raouafi}, {Patsourakos}, {Pariat}, {Young}, {Sterling}, {Savcheva}, {Shimojo}, {Moreno-Insertis}, {DeVore}, {Archontis}, {T{\"o}r{\"o}k}, {Mason}, {Curdt}, {Meyer}, {Dalmasse}, \& {Matsui}}]{Raouafi2016}
{Raouafi}, N.~E., {Patsourakos}, S., {Pariat}, E., {et~al.} 2016, \ssr, 201, 1, \dodoi{10.1007/s11214-016-0260-5}

\bibitem[{{Raouafi} {et~al.}(2023){Raouafi}, {Stenborg}, {Seaton}, {Wang}, {Wang}, {DeForest}, {Bale}, {Drake}, {Uritsky}, {Karpen}, {DeVore}, {Sterling}, {Horbury}, {Harra}, {Bourouaine}, {Kasper}, {Kumar}, {Phan}, \& {Velli}}]{Raouafi2023ApJ...945...28R}
{Raouafi}, N.~E., {Stenborg}, G., {Seaton}, D.~B., {et~al.} 2023, \apj, 945, 28, \dodoi{10.3847/1538-4357/acaf6c}

\bibitem[{{Rochus} {et~al.}(2020){Rochus}, {Auchère}, {Berghmans}, {et~al.}}]{Rochus_2020}
{Rochus}, P., {Auchère}, F., {Berghmans}, D., {et~al.} 2020, \aap, 642, A8

\bibitem[{{Rouppe van der Voort} {et~al.}(2009){Rouppe van der Voort}, {Leenaarts}, {de Pontieu}, {Carlsson}, \& {Vissers}}]{Rouppe+etal2009ApJ...705..272R}
{Rouppe van der Voort}, L., {Leenaarts}, J., {de Pontieu}, B., {Carlsson}, M., \& {Vissers}, G. 2009, \apj, 705, 272, \dodoi{10.1088/0004-637X/705/1/272}

\bibitem[{{Samanta} {et~al.}(2019){Samanta}, {Tian}, {Yurchyshyn}, {Peter}, {Cao}, {Sterling}, {Erd{\'e}lyi}, {Ahn}, {Feng}, {Utz}, {Banerjee}, \& {Chen}}]{Samanta2019}
{Samanta}, T., {Tian}, H., {Yurchyshyn}, V., {et~al.} 2019, Science, 366, 890, \dodoi{10.1126/science.aaw2796}

\bibitem[{{Shen}(2021)}]{Shen_2021}
{Shen}, Y. 2021, Proceedings of The Royal Society A, 477, Article ID 20200217

\bibitem[{{Shi} {et~al.}(2024){Shi}, {Li}, {Ning}, {Xu}, {Song}, \& {Yang}}]{ShiF2024}
{Shi}, F., {Li}, D., {Ning}, Z., {et~al.} 2024, \aap, 686, A279, \dodoi{10.1051/0004-6361/202449377}

\bibitem[{{Shibata} {et~al.}(1992){Shibata}, {Ishido}, {Acton}, {Strong}, {Hirayama}, {Uchida}, {McAllister}, {Matsumoto}, {Tsuneta}, {Shimizu}, {Hara}, {Sakurai}, {Ichimoto}, {Nishino}, \& {Ogawara}}]{shibata92}
{Shibata}, K., {Ishido}, Y., {Acton}, L.~W., {et~al.} 1992, \pasj, 44, L173

\bibitem[{{Sterling} {et~al.}(2016){Sterling}, {Moore}, {Falconer}, {Panesar}, {Akiyama}, {Yashiro}, \& {Gopalswamy}}]{Sterling2016}
{Sterling}, A.~C., {Moore}, R.~L., {Falconer}, D.~A., {et~al.} 2016, \apj, 821, 100, \dodoi{10.3847/0004-637X/821/2/100}

\bibitem[{{Sterling} {et~al.}(2024){Sterling}, {Panesar}, \& {Moore}}]{Sterling2024}
{Sterling}, A.~C., {Panesar}, N.~K., \& {Moore}, R.~L. 2024, \apj, 963, 4, \dodoi{10.3847/1538-4357/ad1d5f}

\bibitem[{{Tandberg-Hanssen}(1977)}]{Tandberg-Hanssen77}
{Tandberg-Hanssen}, E. 1977, Astrophysics and Space Science Library, Vol.~69, {Prominences}, ed. A.~{Bruzek} \& C.~J. {Durrant}, 97

\bibitem[{{Tian} {et~al.}(2014){Tian}, {Kleint}, {Peter}, {Weber}, {Testa}, {DeLuca}, {Golub}, \& {Schanche}}]{Tian+etal.BD.2014ApJ...790L..29T}
{Tian}, H., {Kleint}, L., {Peter}, H., {et~al.} 2014, \apjl, 790, L29, \dodoi{10.1088/2041-8205/790/2/L29}

\bibitem[{{Uitenbroek}(2003)}]{Uitenbroek2003}
{Uitenbroek}, H. 2003, \apj, 592, 1225, \dodoi{10.1086/375736}

\bibitem[{{Yamauchi} {et~al.}(2004){Yamauchi}, {Moore}, {Suess}, {Wang}, \& {Sakurai}}]{Yamauchi2004}
{Yamauchi}, Y., {Moore}, R.~L., {Suess}, S.~T., {Wang}, H., \& {Sakurai}, T. 2004, \apj, 605, 511, \dodoi{10.1086/381240}

\bibitem[{{Yamauchi} {et~al.}(2005){Yamauchi}, {Wang}, {Jiang}, {Schwadron}, \& {Moore}}]{Yamauchi2005}
{Yamauchi}, Y., {Wang}, H., {Jiang}, Y., {Schwadron}, N., \& {Moore}, R.~L. 2005, \apj, 629, 572, \dodoi{10.1086/431664}

\bibitem[{{Yang} {et~al.}(2022){Yang}, {Cao}, \& {Yurchyshyn}}]{2022XYang}
{Yang}, X., {Cao}, W., \& {Yurchyshyn}, V. 2022, \apjs, 262, 55, \dodoi{10.3847/1538-4365/ac91c9}

\bibitem[{{Youn} {et~al.}(2025){Youn}, {Lee}, {Jeong}, {Lee}, {Park}, \& {Moon}}]{youn2025can}
{Youn}, J., {Lee}, H., {Jeong}, H.-J., {et~al.} 2025, \aap, 695, A125, \dodoi{10.1051/0004-6361/202452304}

\bibitem[{{Zhang} \& {Ji}(2014)}]{Zhang2014}
{Zhang}, Q.~M., \& {Ji}, H.~S. 2014, \aap, 567, A11, \dodoi{10.1051/0004-6361/201423698}

\end{thebibliography}
\end{document}